\documentclass[%
 reprint,
superscriptaddress,
 amsmath,amssymb,
 aps,
]{revtex4-2}
\usepackage[version=4]{mhchem}
\usepackage{soul}
\usepackage{siunitx}
\usepackage{color}
\usepackage{graphicx}
\usepackage{dcolumn}
\usepackage{bm}
\usepackage{hyperref}
\usepackage{comment}
\hypersetup{colorlinks=true,linkcolor=red,citecolor=blue}
\usepackage{xr}

\hypersetup{
    colorlinks=true,
    linkcolor=black,    
    citecolor=black,    
    filecolor=black,    
    urlcolor=black      
}

\begin{document}

\preprint{APS/123-QED}
\title{A Magnon Band Analysis of \ce{GdRu2Si2} in the Field-Polarized State}

\author{G. D. A. Wood}
\email{george.wood@stfc.ac.uk}
\affiliation{Department of Physics, University of Warwick, Coventry, CV4 7AL, United Kingdom}
\affiliation{ISIS Facility, STFC Rutherford Appleton Laboratory, Harwell Science and Innovation Campus, Oxfordshire OX11 0QX, United Kingdom}
\author{J. R. Stewart}
\affiliation{ISIS Facility, STFC Rutherford Appleton Laboratory, Harwell Science and Innovation Campus, Oxfordshire OX11 0QX, United Kingdom}
\author{D. A. Mayoh}
\affiliation{Department of Physics, University of Warwick, Coventry, CV4 7AL, United Kingdom}
\author{J. A. M.  Paddison}
\affiliation{Neutron Scattering Division, Oak Ridge National Laboratory, Oak Ridge, TN 37831, USA}
\author{J. Bouaziz}
\affiliation{Peter Gr\"unberg Institut and Institute for Advanced Simulation, Forschungszentrum J\"ulich \& JARA, D-52425 J\"ulich, Germany}
\author{S. M. Tobin}
\affiliation{Department of Physics, University of Oxford, Clarendon Laboratory, Oxford OX1 3PU, United Kingdom}
\affiliation{Australian Nuclear Science and Technology Organisation, Lucas Heights, New South Wales 2234, Australia}
\author{O. A. Petrenko}
\affiliation{Department of Physics, University of Warwick, Coventry, CV4 7AL, United Kingdom}
\author{M. R. Lees}
\affiliation{Department of Physics, University of Warwick, Coventry, CV4 7AL, United Kingdom}
\author{P. Manuel}
\affiliation{ISIS Facility, STFC Rutherford Appleton Laboratory, Harwell Science and Innovation Campus, Oxfordshire OX11 0QX, United Kingdom}
\author{J. B. Staunton}
\affiliation{Department of Physics, University of Warwick, Coventry, CV4 7AL, United Kingdom}
\author{G. Balakrishnan}
\email{g.balakrishnan@warwick.ac.uk}
\affiliation{Department of Physics, University of Warwick, Coventry, CV4 7AL, United Kingdom}

\date{\today}

\begin{abstract}
Understanding the formation of skyrmions in centrosymmetric materials is a problem of fundamental and technological interest. \ce{GdRu2Si2} is one such candidate material which has been shown to host a variety of multi-Q magnetic structures, including in zero-field. Here, inelastic neutron scattering is used to measure the spin excitations in the field-polarized phase of \ce{GdRu2Si2}. Linear spin wave theory and a method of interaction invariant path analysis are used to derive a Hamiltonian accounting for the observed spectra, and comparisons to \textit{ab initio} calculations are made. No evidence for anisotropic or higher order-exchange terms beyond bilinear Heisenberg exchange is found. This is discussed in the context of the multi-Q states existing at lower fields, for which these types of terms have previously been conceived to be significant in the formation of multi-Q ground states.
\end{abstract}

\maketitle

\emph{Introduction} -- Materials with topologically protected properties are expected to prevail in the next generation of electronics\,\cite{he2022topological}. Magnetic skyrmions are a prominent area of research within this field as these vortexlike spin configurations feature remarkable properties including field-driven skyrmion motion, the topological Hall effect and non-reciprocal response which may find applications in spintronics\,\cite{fert2017magnetic,lancaster2019skyrmions,tokura2020magnetic}. Understanding how skyrmions may form at ambient conditions is a key challenge in the drive to discover technologically significant materials. In this regard, centrosymmetric systems are an important class of materials as in principle it has been shown that skyrmions may be stabilized in zero-field\,\cite{wang2021meron,hayami2022zero}, and may also exhibit novel properties such as non-fixed helicity\,\cite{yu2012magnetic,yao2020controlling,yao2022vector}. 

A number of centrosymmetric skyrmion candidates based on \ce{Gd^{3+}} and \ce{Eu^{2+}} intermetallics have been experimentally identified; magnetic ions with weak anisotropy are an important attribute for skyrmion spin-textures. These materials include \ce{Gd2PdSi3}\,\cite{kurumaji2019skyrmion,paddison2022magnetic,ju2023polarized}, \ce{Gd3Ru4Al12}\,\cite{hirschberger2019skyrmion}, \ce{GdRu2Si2}\,\cite{khanh2020nanometric,yasui2020imaging,utesov2021thermodynamically,khanh2022zoology,wood2023double,spethmann2024sp,huddart2024field}, \ce{GdRu2Ge2}\,\cite{yoshimochi2024multistep}, \ce{EuAl4}\,\cite{takagi2022square,hayami2022orthorhombic,gen2023rhombic} and \ce{EuAl_{x}Ga_{2-x}}\,\cite{moya2022incommensurate,PhysRevB.108.L100404}. The latter four systems form in the widely studied \ce{ThCr2Si2}-type parent structure. Aside from the diverse range of emergent physics originating from this structure-type\,\cite{shatruk2019thcr2si2}, the broad chemical tunability of this simple structure makes appealing grounds for theoretical and material discovery initiatives\,\cite{nomoto2023ab}. Putatively, in these materials, the Fermi surface topography and in turn long-ranged Ruderman-Kittel-Kasuya-Yosida (RKKY) interactions act to form the multi-Q spin structures\,\cite{bouaziz2022fermi}. However, there are several differences between the phase diagrams and magnetic structures of each material. This is most clear in the case of \ce{GdRu2Si2} which, unlike the other candidates, has a multi-Q incommensurate ground state in zero field, which is a highly unusual property\,\cite{wood2023double,andriushin2024reentrant,puphal2020topological}. Hence, understanding how the ground state of \ce{GdRu2Si2} forms may provide evidence as to how zero-field skyrmions may be realized. For systems with one magnetic site per primitive unit cell, a key implication of zero-field multi-Q states is the inclusion of anisotropic or higher-order exchange terms in the Hamiltonian (such as biquadratic or four-spin interactions), as the Luttinger-Tisza method stipulates that models with only bilinear Heisenberg exchange terms stabilize single-Q structures\,\cite{luttinger1946theory,lyons1960method,kurz2001three}. A second unresolved aspect of magnetism in \ce{GdRu2Si2} is the large degree of anisotropy between \textit{H-T} phase diagrams for different directions of applied field\,\cite{garnier1996giant, khanh2022zoology}. Naturally, this would imply single-ion anisotropy or anisotropic bilinear exchange interactions entering the Hamiltonian. However, these terms ordinarily arise due to spin-orbit coupling, which ostensibly in an $S=7/2$, $L=0$ system is not present.

In view of gauging the magnetic interactions in \ce{GdRu2Si2}, inelastic neutron scattering measurements of excitations frequently provide stringent and comprehensive tests of magnetic models. However, the application of this method to incommensurate order is in general limited; the magnetic Brillouin zone is very confined and there are a large number of excitation bands, so that there are many closely spaced excitations beyond the resolution capabilities of modern spectrometers\,\cite{janoschek2010helimagnon,kugler2015band,portnichenko2016magnon}. Nevertheless, studying various magnets in the simplified case of the field-polarized phase frequently proves to be an effective method to gain insights into the key magnetic interactions of a system\,\cite{balz2017magnetic,paddison2017continuous}.

In this paper, the spin waves of \ce{GdRu2Si2} in the field-polarized state are examined using inelastic neutron scattering. Analysis of the magnon band is implemented using linear spin wave theory and a method of interaction invariant path analysis. The spectra are accounted for using a set of bilinear Heisenberg exchange interactions up to eighth nearest neighbor, indicative of the long-range RKKY interactions. Qualitatively, the model agrees well with \textit{ab initio} calculations\,\cite{bouaziz2022fermi}. No evidence for higher-order or anisotropic exchange interactions is found. The implications of the derived Hamiltonian are discussed within the context of the Fermi surface reconstruction that occurs between the field-polarized and the lower-field multi-Q states\cite{matsuyama2023quantum}.

\begin{figure}[tb]
\includegraphics*[width=1\linewidth,clip]{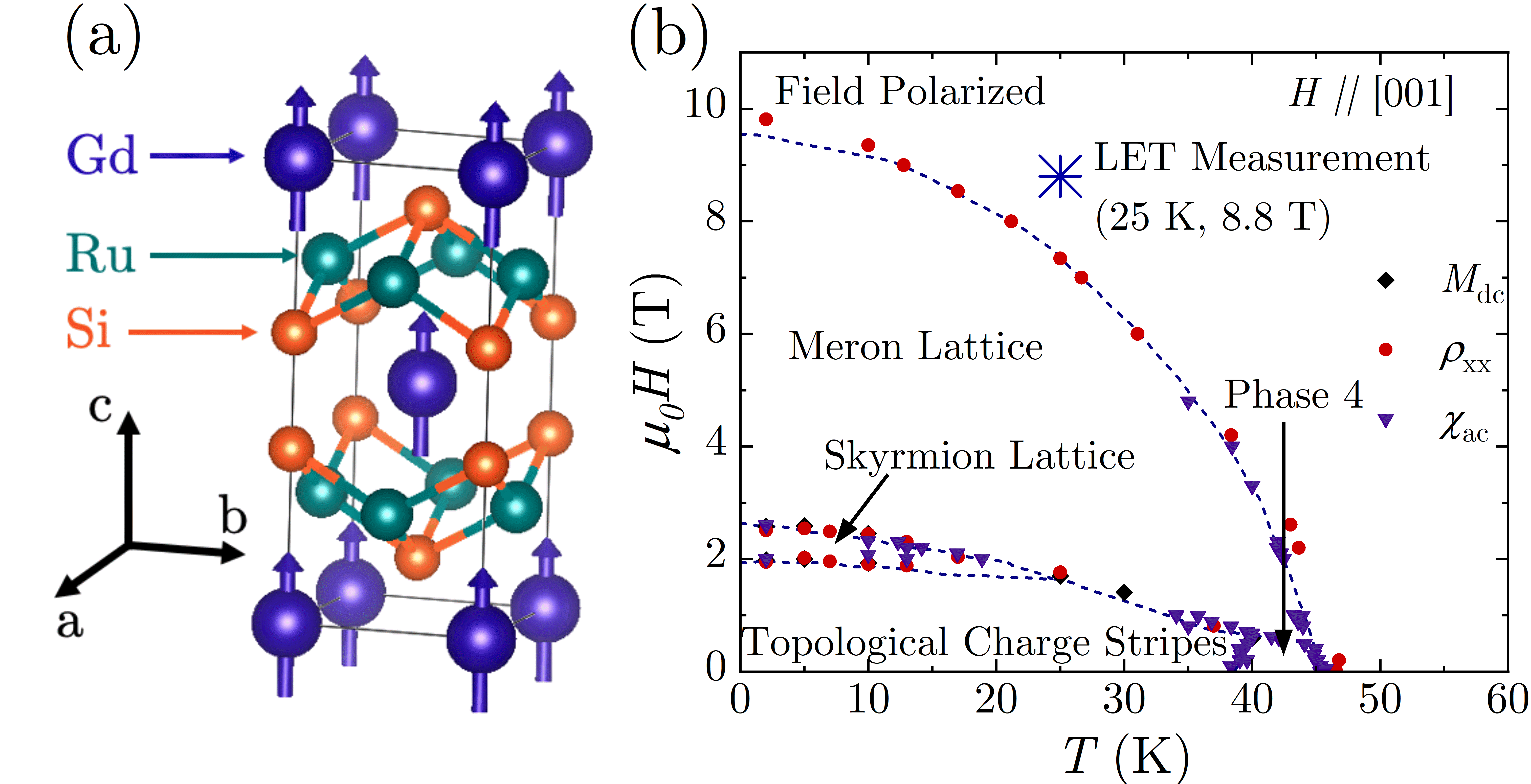}
\caption{\label{fig:HTPD}(a) Conventional unit cell of \ce{GdRu2Si2}, which forms in the $I4/mmm$ space group (with $a=4.16\,\si{\angstrom}$ and $c=9.59\,\si{\angstrom}$) with the Gd moments depicted in a field-polarized state. (b) $H$-$T$ phase diagram of \ce{GdRu2Si2} with $H$ applied along the $c$\,axis. The position of the LET \textit{Horace} scan is marked at (25\,K, 8.8\,T).}
\end{figure}

\emph{Experimental Details} -- To enable neutron scattering $98.1\%$ isotopically enriched \ce{^{160}Gd} was used\,\cite{Gd}. The \ce{^{160}GdRu2Si2} single crystal boule, $17\,\si{mm}$ in length and $\sim2.3\,\si{mm}$ in diameter, the same as that used in Ref.\,\cite{wood2023double}, was mounted on an aluminum strip holder. Single-crystal inelastic neutron scattering measurements were performed on the LET spectrometer (a direct geometry time-of-flight instrument) at the ISIS Neutron and Muon Source\,\cite{bewley2011let}.  Measurements were taken at 25\,K with a 8.8\,T magnetic field applied along the $c$~axis so that the sample was in the forced ferromagnetic state (see Fig.\,\ref{fig:HTPD}) with $a^{*}$-$b^{*}$ as the primary scattering plane. A principal incident energy of $E_{i}=7.52\,\si{meV}$ was used with a percentage resolution of $3.7\%$ at the elastic line~\cite{LETE}. Note that despite the \ce{^{160}Gd} enrichment, the sample is very absorbing with $\bar{\sigma_{a}}(E_{i}=7.52\,\si{meV})\approx1736\,\si{b}$ and hence the signal-to-noise ratio of the measurement throughout
is low. A $90^{\circ}$ \textit{Horace} scan was taken with $0.5^{\circ}$ steps and $\sim13$ minutes of counting per step ($\sim8\,\si{\micro\ampere\hour}$). The \textit{Mantid} software was used for data normalization and the \textit{Horace} package was used to construct four-dimensional $S(\mathbf{Q},\omega)$ datasets\,\cite{arnold2014mantid,ewings2016horace}.

\begin{figure*}
    \centering
    \includegraphics[width=\textwidth]{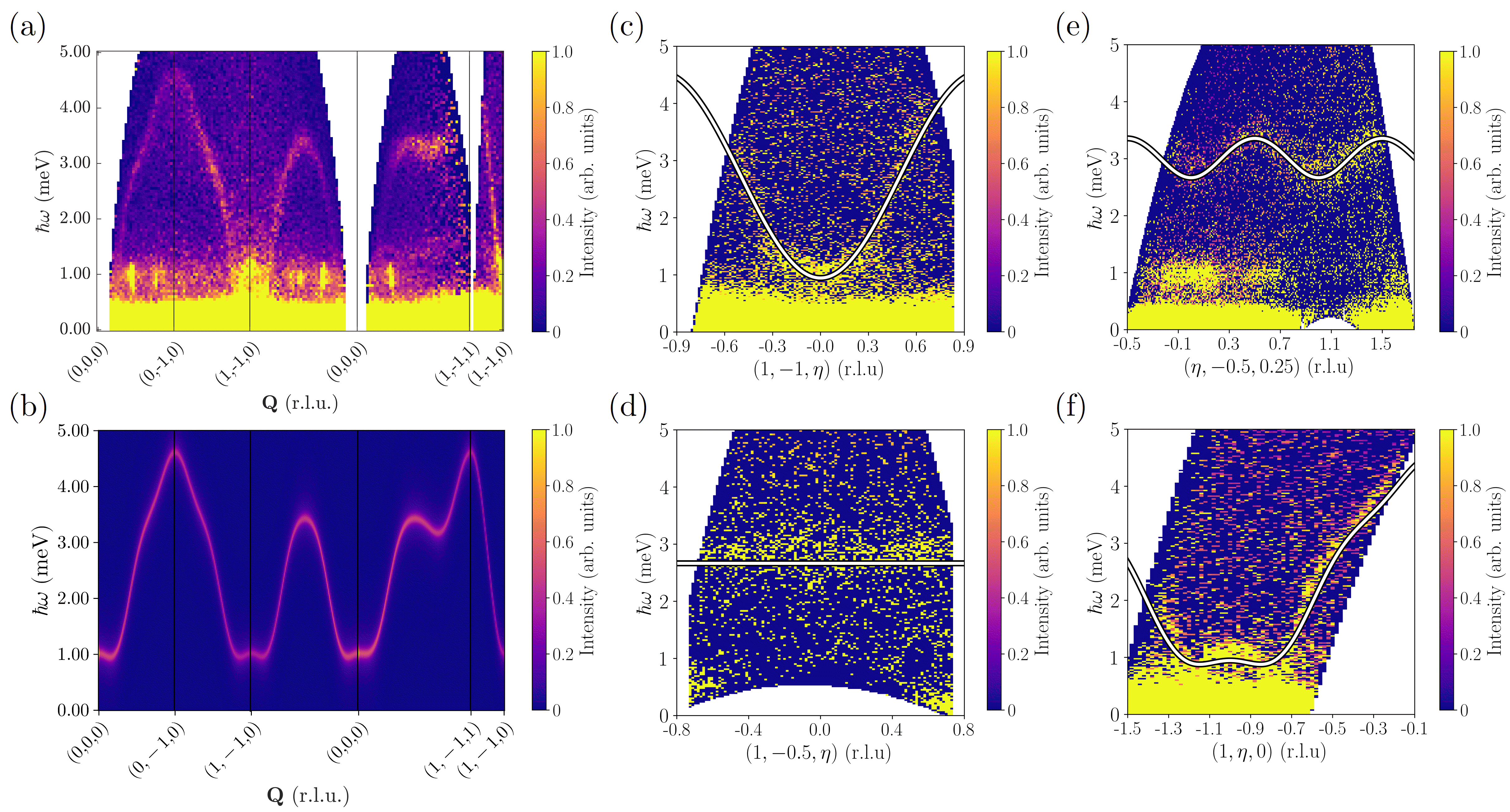}
    \caption{Various momentum-resolved inelastic neutron scattering spectra of \ce{GdRu2Si2} at $T=25$\,K and $B=8.8$\,T in the field-polarized phase ($E_{i}=7.52\,\si{meV}$). A single magnon band is observed throughout. (a) An overview of the inelastic neutron scattering data . (b) Corresponding calculation of the magnon dispersion model implemented in the \textit{Sun(n)y} code\,\cite{zhang2021classical,dahlbom2022geometric}. (c) An out-of-plane magnon dispersion along the $(1,-1,\eta)$ path. $J_{2}$ is active and dominant along this cut. (d) A $J_{2}$ invariant cut on the $(1,-0.5,\eta)$ path; a flat band is observed and this indicates $J_{7}$ is negligible. (e) A $J_{2}$ invariant cut along the $(\eta,-0.5,0.25)$ path; the magnon band is dispersive and therefore several in-plane interactions are required in the model. (f) A cut along $(1,\eta,0)$, which passes directly over a zone center $(1,-1,0)$. The `\textit{w}' shape at the bottom of the dispersion curve is indicative of the incommensurate order stabilized at lower fields. In (c)-(f) the white/black line superimposed on the data is calculated using the spin-wave model described in the text. The complete set of cuts used for fitting the linear spin wave model is shown in the Supplemental Material (Figs.\,S2, S3 and S4)\,\cite{SI}.}
    \label{fig: Magnons}
\end{figure*}

\emph{Analysis} -- An overview of the inelastic neutron spectra is provided in Fig.\,\ref{fig: Magnons}. The observation of a single band is consistent with \ce{Gd^{3+}} being the only magnetic ion within the primitive unit cell. In addition to the magnon band there are several low energy spurious signals originating from the sample environment (\textit{e.g.} Figs.\,\ref{fig: Magnons}(a) and \ref{fig: Magnons}(e) at $\hbar\omega\sim\,0.8\,\si{meV}$), which, since the excitation signal from the sample is intrinsically low due to the absorption, appear comparatively strong. So that the magnon band signal could be systematically extracted, median kernel filtering was applied; this methodology is described in detail in the Supplemental Material\,\cite{SI}.

To model the magnon band we assume a Hamiltonian with bilinear Heisenberg exchange, Zeeman coupling and single-ion anisotropy,
\begin{equation}
H=\sum_{\left \langle i,j  \right \rangle}\left\{J_{i,j}\mathbf{S}_{i}\cdot\mathbf{S}_{j} + g\mu_{\mathrm{B}}\mathbf{B}\cdot\mathbf{S}_{i}+K(S_{i}^{z})^{2}\right\}.
\label{Hamiltonian}
\end{equation}
\noindent Applying the Holstein-Primakoff transformation provides the linear spin-wave dispersion for a ferromagnet,
\begin{equation}
\hbar\omega(\mathbf{q}) = S[J(\mathbf{q}) - J(\mathbf{0})] - g\mu_{\mathrm{B}}B-2KS,
\label{lswt}
\end{equation}
\noindent where $J(\mathbf{q})$ is defined as,
\begin{equation}
J(\mathbf{q}) = \sum_{\boldsymbol{\delta}\neq 0 }J_{\boldsymbol{\delta}}\exp(i\mathbf{q}\cdot\boldsymbol{\delta}).
\label{Jofq}
\end{equation}
\noindent Here, $\boldsymbol{\delta}$ is the interaction vector ($\mathbf{r}_{i}-\mathbf{r}_{j}$) between the magnetic ions expressed in lattice units and $\mathbf{q}$ is a reciprocal space vector (in r.l.u.)\,\cite{BCC, boothroyd2020principles}. Initial attempts at fitting the magnon dispersion revealed that $J_{2}$ (with $\boldsymbol{\delta}=[1/2,1/2,1/2]$) is the dominant interaction, but that multiple interactions were required to account for the dispersion character, consistent with long-ranged RKKY interactions. To gauge which interactions are active in the system we have developed a method of interaction invariant path analysis. In this technique, reciprocal space cuts of $S(\mathbf{Q},\omega)$ are chosen so that a specified interaction is inactive; \textit{i.e.} the spin-wave theory pertaining to the interaction is non-dispersive along the reciprocal-space path. By isolating certain interactions, weaker, but nevertheless significant exchange interactions may be refined. In particular we make use of $J_{2}$ invariant paths. The linear spin wave theory for the $J_{2}$ interaction is given by,
\begin{equation}
\hbar\omega(\mathbf{q})= 8SJ_{2}\bigl\{\cos(\pi q_{x})\cos(\pi q_{y})\cos(\pi q_{z}) - 1\bigr\}. 
\label{j2q}
\end{equation}

\noindent Cancellation of the cosine terms in Eq.\,\ref{j2q} is provided when $q_{x}, q_{y}$ or $q_{z} = 1/2$ so that the $J_{2}$ term will only manifest as a constant in the dispersion relation. For instance, consider the $(1,-1,\eta)$ cut shown in Fig. \ref{fig: Magnons} (c). A strong parabolic dispersion dominated by the $J_{2}$ interaction is evident. Conversely, the $(1,-0.5,\eta)$ cut in Fig. \ref{fig: Magnons} (d) is a $J_{2}$ invariant cut with $q_{y}=-1/2$. This invariant path is effective for constraining out-of-plane interactions, such as $J_{7}$ ($\boldsymbol{\delta}=[0,0,1]$). Since a flat band is observed this implies that $J_{7}$ is negligible, and therefore it is not included as a parameter in the model. Similarly, the $(\eta,-0.5,0.25)$ cut of Fig. \ref{fig: Magnons} (e) is a $J_{2}$ invariant, and this path is effective for constraining interactions  with $\boldsymbol{\delta}=[r_{x},r_{y},r_{z}=0]$, which includes $J_{1}$ and $J_{3}$ (see Table \ref{table:GRSinteractions}). Since the magnon band along this path is dispersive, these parameters are included in the model. Additional notes for this new methodology are provided in the Supplemental Material\,\cite{SI}.

\begin{table}
\centering
\begin{tabular}{||c c c||} 
 \hline
 Interaction (\si{\micro eV}) & $\boldsymbol{\delta}$ & Multiplicity \\ [0.5ex] 
 \hline\hline
 $J_{1}=-33.9\pm1.0$ & $[1,0,0]$ & 4 \\ 
 $J_{2} = -65.1\pm0.7$ & $[1/2,1/2,1/2]$ & 8  \\
 $J_{3} = 13.6\pm0.7$ & $[1,1,0]$ & 4  \\
 $J_{4} = -1.1\pm0.4$ & $[3/2,1/2,1/2]$ & 16  \\
 $J_{5} = 10.8\pm0.8$ & $[2,0,0]$ & 4 \\
 $J_{6} = 6.0\pm0.5$  & $[1,2,0]$ & 8  \\
 $J_{7} = 0$ & $[0,0,1]$ & 2 \\
 $J_{8} = 3.0\pm0.4$ & $[3/2,3/2,1/2]$ & 8 \\[1ex]
 \hline
\end{tabular}
\caption{A list of bilinear Heisenberg interactions and the experimentally derived interaction energies of the $J_{1\rightarrow 8}$ model. $\boldsymbol{\delta}$ is the vector in lattice units over which each interaction acts (not including symmetry equivalents). The multiplicity of the interaction counts symmetry equivalent $\boldsymbol{\delta}$ vectors for each interaction.}
\label{table:GRSinteractions}
\end{table}

The measurement did not contain sufficient intensity to reliably fit $S(\mathbf{Q},\omega)$, hence an $\hbar\omega(\mathbf{q})$ fitting was implemented instead using a least-squares minimization routine with the \textit{LMFIT} package\,\cite{newville2016lmfit}. In total, seven bilinear Heisenberg exchange interactions and a constant parameter ($C$) were used to fit the magnon band (see table \ref{table:GRSinteractions})\,\cite{CutTB}. Only including $J_{1}$ to $J_{5}$ resulted in dispersions with global minima coinciding with zone centers which would be consistent with an intrinsic ferromagnet. With a $J_{1}$ to $J_{6}$ model the dispersion relation adopted global minima surrounding the zone center at incommensurate positions, but these minima were along the $[1,1,0]$ direction which would be inconsistent with the direction of the magnetic propagation in the lower-field phases. By including a $J_{8}$ interaction the dispersion relation adopts global minima at $\mathbf{q}=[0.15,0,0]$ which is the correct direction for magnetic propagation. Conceivably, if the magnetic field was lowered, then the magnon band would lower and where the minima coincide with $\hbar\omega=0$ a field-polarized$\rightarrow$incommensurate phase transition would take place. Models including further exchange interactions were found to marginally increase the quality of the fit; however, this came at the expense of higher correlation between parameters.  With a preference for a minimal effective model that captures the salient features of the spin-wave dispersion, the $J_{1}$ to $J_{8}$  ($J_{1\rightarrow 8}$) model suffices. The constant term in the model refined to $C=0.94\pm0.01\,\si{meV}$. Noting that $-g\mu_{B}B=1.02\,\si{meV}$, we consider the refined value $C$ to be broadly consistent with $-g\mu_{B}B$ when demagnetization effects would be expected to slightly lower the contribution of the field. Implicitly, from Eq. \ref{lswt}, it is not possible to disentangle constant contributions to the dispersion relation \textit{i.e.} the $-g\mu_{B}B$, $-2KS$, and any unaccounted interaction terms. Indeed, depending on the number of exchange terms included in the model, the refined value of $C$ varied between $0.88\sim0.99\,\si{meV}$ generally increasing with the number of exchange terms included. Therefore we attribute the difference $\delta=-g\mu_{B}B-C$ as likely arising from small unaccounted exchange terms, and consider $C$ to be consistent with that of the applied magnetic field with negligible single-ion anisotropy in the Hamiltonian.

We also evaluated the effect of dipolar interactions on the linear spin-wave model. Dipolar interactions are an intrinsic source of anisotropy since these interactions are directly coupled to the lattice geometry and therefore could be important in explaining the anisotropic phase diagrams and multi-Q spin textures existing at lower fields.  These calculations were implemented with Ewald summation using the \textit{Sun(n)y} code\,\cite{zhang2021classical,dahlbom2022geometric} and are presented in Fig.\,S6 of the Supplemental Material\,\cite{SI}. We find that the dipolar interaction has a very small effect on the spin wave calculations, only affecting the dispersion curve at the top and bottom of the band where the difference between the bands of the original and dipolar-included model are imperceptible in terms of the measurement.

\begin{figure}[tb]
\includegraphics*[width=0.8\linewidth,clip]{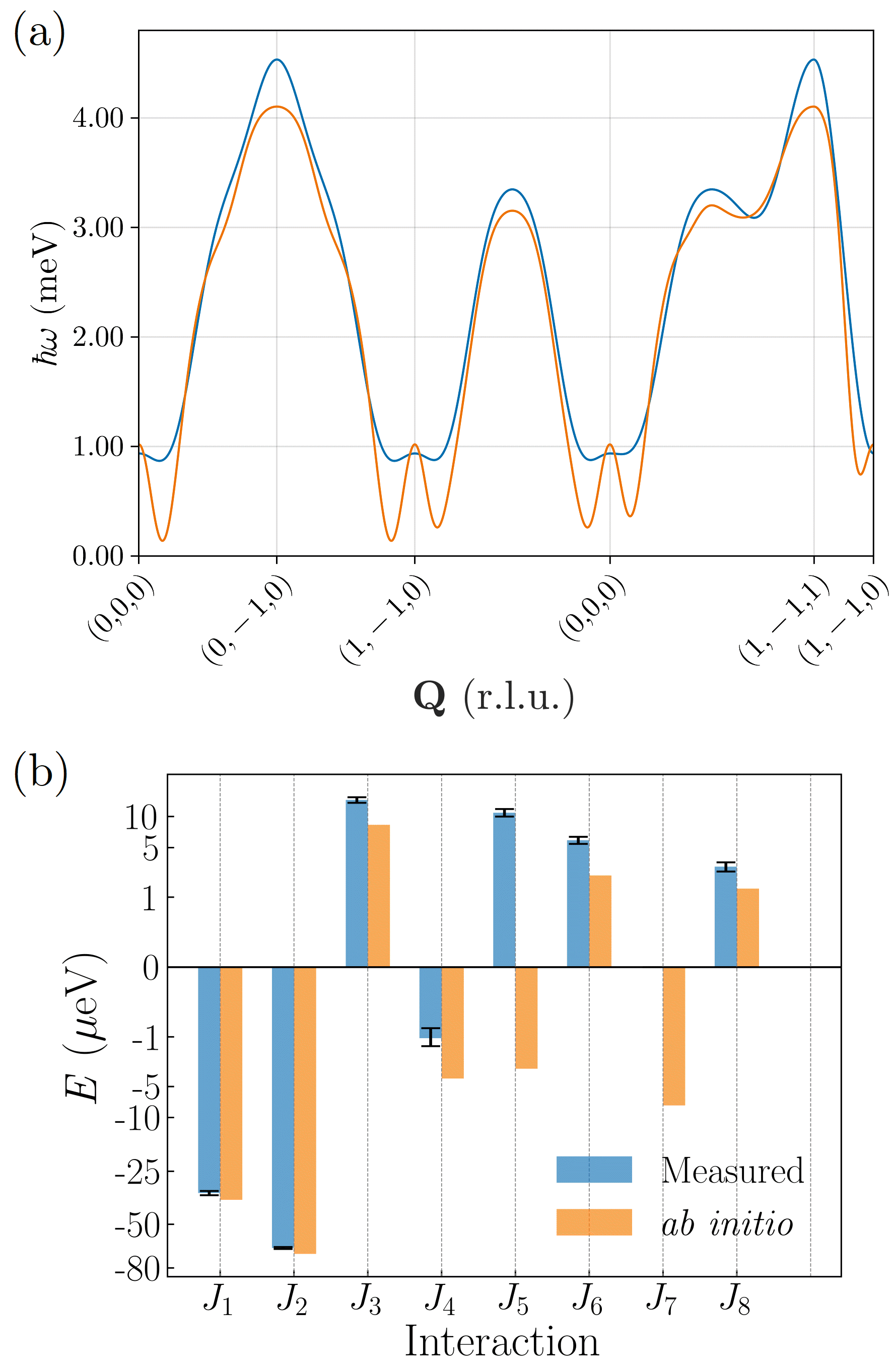}
\caption{\label{fig: Comparison}(a) Comparison of dispersions between the model derived by Bouaziz \textit{et al.,}\,\cite{bouaziz2022fermi} (using $J_{1}$ to $J_{50}$) represented by the orange line and the $J_{1\rightarrow 8}$ effective model represented by the blue line. (b) Comparison of the experimentally determined Heisenberg exchange terms in \ce{GdRu2Si2} with those theoretically derived by Bouaziz \textit{et al.,}\,\cite{bouaziz2022fermi}. Note that a cube root scale is used along the $y$~axis for approximate RKKY scaling ($J_{RKKY}\sim r^{-3}$).}
\end{figure}

\emph{Discussion} -- A comparison between the \textit{ab initio} model derived by Bouaziz \textit{et al.,}\,\cite{bouaziz2022fermi} and the $J_{1\rightarrow 8}$ model is shown in Fig. \ref{fig: Comparison}. In most aspects it is clear that both models are qualitatively consistent and there is a remarkable likeness between the dispersion shapes which substantiates the veracity of the \textit{ab initio} calculations by Bouaziz \textit{et al.,}\,\cite{bouaziz2022fermi}. This conclusion is consistent with the recent zero-field powder study of \ce{GdRu2Si2} by Paddison \textit{et al.,}\,\cite{paddison2024spin}. From Fig. \ref{fig: Comparison} (a) the two main points of difference between the models are at the top and bottom of the dispersions. The maxima of the dispersion model of Bouaziz \textit{et al.,} underestimates the measured magnon band and does not fit the data. However, at the bottom of the band, the model of Bouaziz \textit{et al.,} is likely more representative of the underlying incommensurate physics than the $J_{1\rightarrow 8}$ model. Firstly the global minimum of the \textit{ab initio} model is at $(0.18,0,0)$ which is closer to the magnetic propagation vector of $\mathbf{q}_{e}=[0.22,0,0]$ as determined by high-resolution diffraction experiments\,\cite{khanh2020nanometric,khanh2022zoology,paddison2024spin}. Likely, the reason the global minimum of the $J_{1\rightarrow 8}$ model is substantially less than $\mathbf{q}_{e}$ is since the bottom of the magnon band is not directly resolvable in the measurement, since the strong tail of elastic scattering obscures these details for $\hbar\omega<0.8\,\si{meV}$ (see Fig. \ref{fig: Magnons} (f)).  Without this information, the determination of the global minimum of the magnon band is somewhat uncertain and the fitted model can only infer these minima from higher energy portions of the measured dispersion relation. Following this, given that at $25\,\si{K}$ the system enters the field-polarized phase at $7.36\,\si{T}$, then with the applied magnetic field of $\,8.8\,\si{T}$ we would estimate that the bottom of the dispersion curve should sit at $-g\mu_{B}\Delta B\approx0.17\,\si{meV}$. Again, the minimum of the dispersion for the model of Bouaziz \textit{et al.,} is closer to this calculation. With regards to the $J_{1\rightarrow 8}$ model we adduce that additional exchange interactions would be required to recover more realistic global minima in the dispersion relation. Indeed, this was a key aspect of the model derived by Bouaziz \textit{et al.,}  which found that a long-range \textit{tail} of RKKY interactions had to extend up to ten unit cells\,\cite{bouaziz2022fermi}; the point being, that it was the cumulative effect of these weak interactions which was central in determining a propagation vector in reasonable agreement with experiment. However, without directly measuring the bottom of the magnon band below $\hbar\omega<0.8\,\si{meV}$ it is not possible to fit more interactions\,\cite{FutureExpts}, and in any case, fitting a long-range \textit{tail} of interactions within ten unit cells would introduce an intractable number of parameters. Supposing that there are unaccounted interactions in the $J_{1\rightarrow 8}$ model, we calculate a mean-field estimation of the Curie-Weiss temperature for the model: $\Theta=-\frac{S(S+1)}{3}J(\mathbf{0})=30.5\,\si{K}$. Comparing this to the experimental value of $42.3\,\si{K}$ (see Fig.\,S7) indicates that our model accounts for $72\%$ of net ferromagnetic interactions in the system. This conclusion is corroborated by Fig.~\ref{fig: Magnons} (d) which shows that the flat-band is slightly underestimated in the spin-wave model. Finally, Fig.~\ref{fig: Comparison}(b) directly compares the values of the exchange terms in both models. There is very good agreement between exchange values for the first few parameters, but there is clear departure between the models for $J_{5}$ and $J_{7}$. The omission of $J_{7}$ in the effective model has been motivated by the flat band in Fig.~\ref{fig: Magnons}(d). The reason that $J_{5}$ is required to be antiferromagnetic in the $J_{1\rightarrow 8}$ model is so that the global minimum of the band lies at incommensurate positions; this particular difference between the \textit{ab initio} model of Bouaziz \textit{et al.,} is likely due to not including the long-range exchange parameters ($J_{n>8}$) in the effective model.

More generally, the model Hamiltonian can be considered in the context of the  lower-field multi-Q states and the anisotropy in the phase diagrams\,\cite{wood2023double,garnier1996giant}. We have determined that the magnon band is dominated by a set of bilinear (RKKY) Heisenberg interactions. Crucially, to model the magnon band, higher-order exchange interactions, such as anisotropic and biquadratic exchange, are not required. Therefore, if these terms do exist, then it is likely they are small and limited to $\lesssim 1\,\si{\mu eV}$.  This is consistent with the calculations of Bouaziz \textit{et al.,}\,\cite{bouaziz2022fermi} which found negligible evidence for these types of interactions. Clearly, though, these aspects are difficult to reconcile with the zero-field multi-Q ground state and a substantial degree of anisotropy in the $H$-$T$ phase diagrams. Frequently, in zero-field multi-Q  magnetic order models, anisotropic and higher-order exchange interactions are central terms in the model Hamiltonian\,\cite{khanh2022zoology,wang2021meron,hayami2024three,hayami2024double,paddison2024cubic,hayami2024stabilization}. This derives as one of the consequences of the Luttinger-Tisza method, that a Hamiltonian containing just bilinear Heisenberg exchange terms stabilizes single-Q helical solutions\,\cite{lyons1960method}. This poses a problem, since this is a system which hosts zero-field multi-Q order, but the excitation spectra and \textit{ab initio} calculations find no evidence for the types of exchange interactions which could stabilize such order. In view of this we put forward the following proposal. 

We posit firstly that a more significant form of anisotropy originates from the RKKY interaction and the Fermi surface topography. The RKKY interaction develops as a feedback effect between valence electrons at the Fermi surface and the order on the localized  $4f$ magnetic moments. Tacitly then, it is a function of both temperature and magnetic field (\textit{i.e.} $J_{i,j}(\mathbf{B},T))$, as through Neumann's principle the $C_{4}$ symmetry of the system will be broken (or preserved) according to the direction of the applied field, which should manifest in the Fermi surface topography and RKKY potential. The notion of an evolving Fermi surface between magnetic phases in this system is consistent with the nesting conditions which have been identified in angle-resolved photoemission spectroscopy studies\,\cite{eremeev2023insight,dong2023magnetic} and the quantum oscillations study of Matsuyama \textit{et al.,} that associated a Fermi surface reconstruction with a change in oscillation frequency in measurements between the field-polarized and lower-field multi-Q state\,\cite{matsuyama2023quantum}. If the RKKY interactions are a function of field and temperature then this may explain why the $J_{1\rightarrow 8}$ model in the field-polarized phase is not entirely compatible with the incommensurate order existing at lower-fields.

Second, in the multi-Q phases of interest there is intricate coupling between magnetic and charge-density wave order parameters, as discovered by Yasui \textit{et al.,}\,\cite{yasui2020imaging}. Currently, the precise details of the symmetry breaking of the charge density waves are not fully understood. Irrespective of these details though, the charge density waves will distort the crystal structure and the incommensurate charge density will act to modulate the Heisenberg exchange network across the system\,\cite{zaliznyak2003heisenberg,zaliznyak2004spiral}. These aspects may give rise to more complex magnetic ordering in zero magnetic field. 

\emph{Conclusion} -- A magnon band in the field-polarized phase of \ce{GdRu2Si2} has been measured using time-of-flight inelastic neutron scattering. A Hamiltonian accounting for the dispersion relation has been derived with linear spin wave theory. The model, using eight free parameters, is in good accordance with \textit{ab initio} calculations\,\cite{bouaziz2022fermi}, and captures the salient features of the magnon dispersion, including global minima at incommensurate positions which are characteristic of the lower-field magnetism. There are a few indications that more exchange interactions are present in the system, with the cumulative effect of long-range RKKY interactions being significant. In principle, these would improve certain aspects of the model; however, without resolution of the dispersion at the lowest energy transfers, this would introduce many parameters to the model which the current dataset does not support. We have discussed the derived model in the context of the lower-field multi-Q magnetism and the evolving Fermi surface as a function of applied field direction and temperature. Future experimental efforts will focus on understanding the coupling between charge density waves and magnetic order to understand the origin of the zero-field ordering.

\emph{Acknowledgments} -- We are grateful for useful discussions with Jacob Wilkins (Scientific Computing Department, STFC), Manh Duc Le, Toby Perring, Travis Williams and David Voneshen (ISIS Neutron and Muon Source, STFC). The work of J.A.M.P. was supported by the U.S. Department of Energy, Office of Science, Basic Energy Sciences, Materials Sciences and Engineering Division. This paper was financially supported by two Engineering and Physical Sciences Research Council grants: EP/T005963/1, the UK Skyrmion Project, EP/N032128/1 and by U.K. Research and Innovation and Science and Technology Facilities Council through the provision of beam time at the ISIS Neutron and Muon Source\,\cite{Gdata1}, as well as partial funding for the \ce{^{160}Gd} isotope used in the experiment.

\bibliography{Main.bib}

\begin{thebibliography}{64}
\bibitem [{LET()}]{LETE}%
  \BibitemOpen
  \href@noop {} {}\bibinfo {note} {With the multiplexing capability of LET, three other incident energies were available ($E_{i}=2.2$, $3.7$ and $22.80\, \si{meV}$). The two lower incident energies would have been useful for resolving the dispersion at the lowest energies. However, the absorption of the sample at these lower energies was more severe and no magnon signal was perceptible.}\BibitemShut {Stop}%
\bibitem [{BCC()}]{BCC}%
  \BibitemOpen
  \href@noop {} {}\bibinfo {note} {Note that throughout the manuscript we retain the conventional body-centered setting of \ce{GdRu2Si2} ({$I4/mmm$}).}\BibitemShut {Stop}%
\bibitem [{Cut()}]{CutTB}%
  \BibitemOpen
  \href@noop {} {}\bibinfo {note} {We define $B=-8.8~\si{T}$ so that it is directed along the $z$ direction in this model.}\BibitemShut {Stop}%
\end{thebibliography}

\end{document}


\title{Supplementary Information: A Magnon Band Analysis of \ce{GdRu2Si2} in the Field-Polarized State}

\author{G. D. A. Wood}
\email{george.wood@stfc.ac.uk}
\affiliation{Department of Physics, University of Warwick, Coventry, CV4 7AL, United Kingdom}
\affiliation{ISIS Facility, STFC Rutherford Appleton Laboratory, Harwell Science and Innovation Campus, Oxfordshire OX11 0QX, United Kingdom}
\author{J. R. Stewart}
\affiliation{ISIS Facility, STFC Rutherford Appleton Laboratory, Harwell Science and Innovation Campus, Oxfordshire OX11 0QX, United Kingdom}
\author{D. A. Mayoh}
\affiliation{Department of Physics, University of Warwick, Coventry, CV4 7AL, United Kingdom}
\author{J. A. M. Paddison}
\affiliation{Neutron Scattering Division, Oak Ridge National Laboratory, Oak Ridge, TN 37831, USA}
\author{J. Bouaziz}
\affiliation{Peter Gr\"unberg Institut and Institute for Advanced Simulation, Forschungszentrum J\"ulich \& JARA, D-52425 J\"ulich, Germany}
\author{S. M. Tobin}
\affiliation{Department of Physics, University of Oxford, Clarendon Laboratory, Oxford OX1 3PU, United Kingdom}
\affiliation{Australian Nuclear Science and Technology Organisation, Lucas Heights, New South Wales 2234, Australia}
\author{O. A. Petrenko}
\affiliation{Department of Physics, University of Warwick, Coventry, CV4 7AL, United Kingdom}
\author{M. R. Lees}
\affiliation{Department of Physics, University of Warwick, Coventry, CV4 7AL, United Kingdom}
\author{P. Manuel}
\affiliation{ISIS Facility, STFC Rutherford Appleton Laboratory, Harwell Science and Innovation Campus, Oxfordshire OX11 0QX, United Kingdom}
\author{J. B. Staunton}
\affiliation{Department of Physics, University of Warwick, Coventry, CV4 7AL, United Kingdom}
\author{G. Balakrishnan}
\email{g.balakrishnan@warwick.ac.uk}
\affiliation{Department of Physics, University of Warwick, Coventry, CV4 7AL, United Kingdom}

\maketitle
\tableofcontents

\newpage
\section{Signal Extraction of the Magnon Band}
\vspace{1em}

As stated in the main manuscript, the signal-to-noise ratio of the magnon band across the LET measurement is low due to the strong absorption of the sample. In this case, directly fitting an $S(\mathbf{Q},\omega)$ model to the dataset was unsuccessful since the $S(\mathbf{Q},\omega)$ fitting software\,\footnote{\textit{Tobyfit} as part of the \textit{Horace} package was attempted for this\,\cite{ewings2016horace}} cannot effectively distinguish between noise and signal. Nevertheless, the magnon band is clearly perceptible in the measurement. Therefore to extract the signal, median kernel filtering was applied as a nonlinear filter to reduce the effects of noise\,\cite{arce2004nonlinear}, and to identify the pixels which contain the spin-wave signal. We have applied a three-step process for this.

\begin{enumerate}
  \item On a given cut, mask the tail of the elastic scattering at low energies and spurious background signals in the measurement. The cuts should be finely binned, so that if there is random noise, the surrounding pixels likely have a zero value. This is essential for the median filter to be effective.
  \item Apply a $3\times3$ median-kernel grid to each pixel in a given cut. After this process, extract any pixels which have non-zero median values, assuming that these pertain to the magnon band. We have used the \textit{NDimage} code from \textit{SciPy} to perform this image processing procedure\,\cite{2020SciPy-NMeth}. 
  \item So that the fitting procedure operates on real data, we map back to the original intensity values of the pixels which identified as non-zero in the median kernel filter process. We use the intensity of the pixels to weight the $\hbar\omega(\mathbf{q)}$ fitting. This is motivated by assuming that the cross-section of the magnon band at a given ($\mathbf{q},\omega)$ coordinate has a Gaussian-like profile when making a perpendicular cut to the dispersion curve.
\end{enumerate}

\begin{figure*}
    \centering
    \includegraphics[width=1\textwidth]{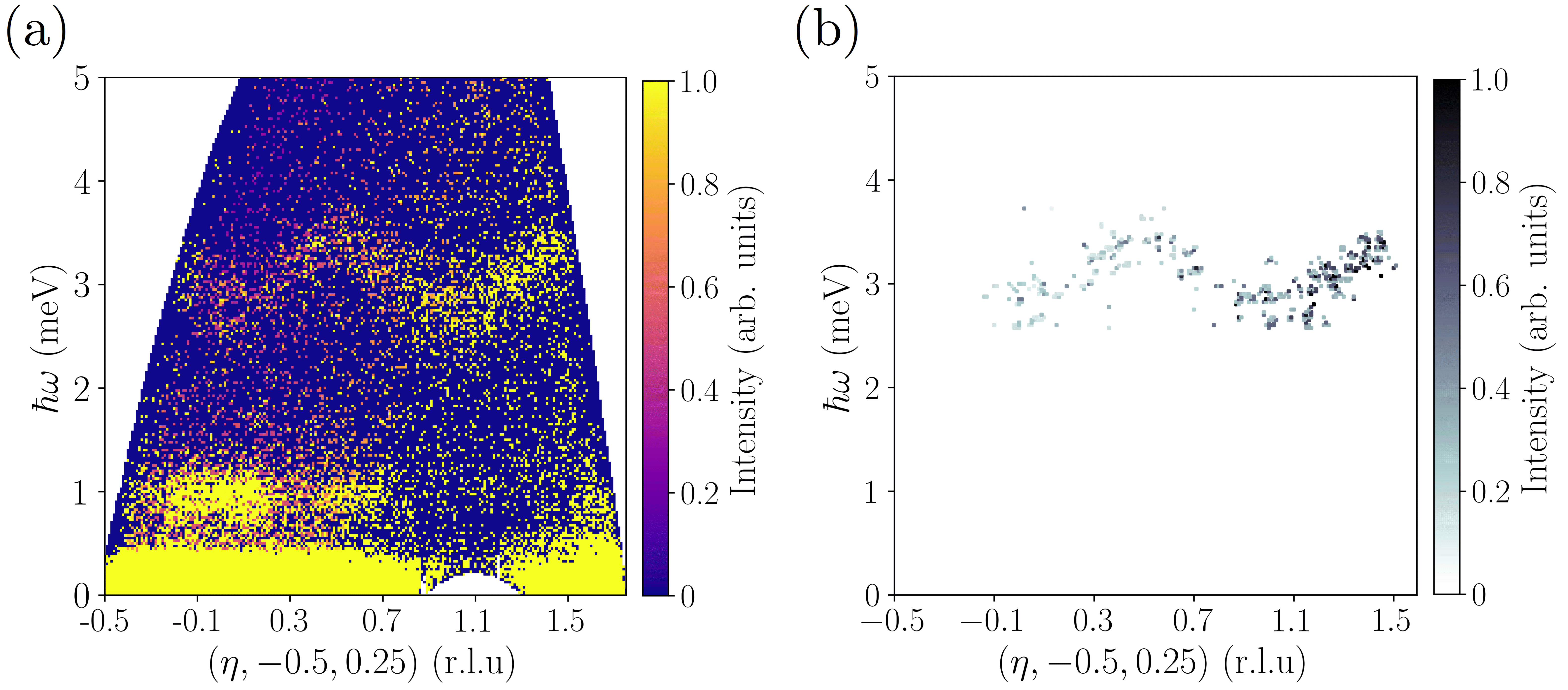}
    \caption{(a) Reciprocal space cut along the $(\eta,-0.5,0.25)$ path before median filter processing. (b) Processed cut after median kernel filtering.}
    \label{fig: SignalExtraction}
\end{figure*}

The results of this process between the original data and the filtered data are shown in Figs.~\ref{fig: SignalExtraction}(a) and \ref{fig: SignalExtraction}(b), which show an example cut before and after this process, respectively. The median filtered results for all cuts used to fit the magnon dispersion are shown in the central column in Figs.\,\ref{fig: MultiBaltis1}, \ref{fig: MultiBaltis2} and \ref{fig: MultiBaltis3}.

Generally, with direct geometry time-of-flight spectrometers, the signal-to-noise ratio is often low when measuring four-dimensional $S(\mathbf{Q},\omega)$ datasets due to the limited neutron flux. Therefore, these non-linear kernel methods may find broader application in the quantitative treatment of such measurements. Here, two-dimensional kernels have been applied, but in principle four-dimensional kernels may be applied to the $(\mathbf{Q},\omega)$ domain for more efficient signal extraction, and plans exist to integrate these features into $S(\mathbf{Q},\omega)$ handling software\,\cite{ewings2016horace}.

\section{Model Fitting and the Complete Set of Reciprocal Space Cuts}

\begin{figure}
    \centering
    \includegraphics[width=1\textwidth]{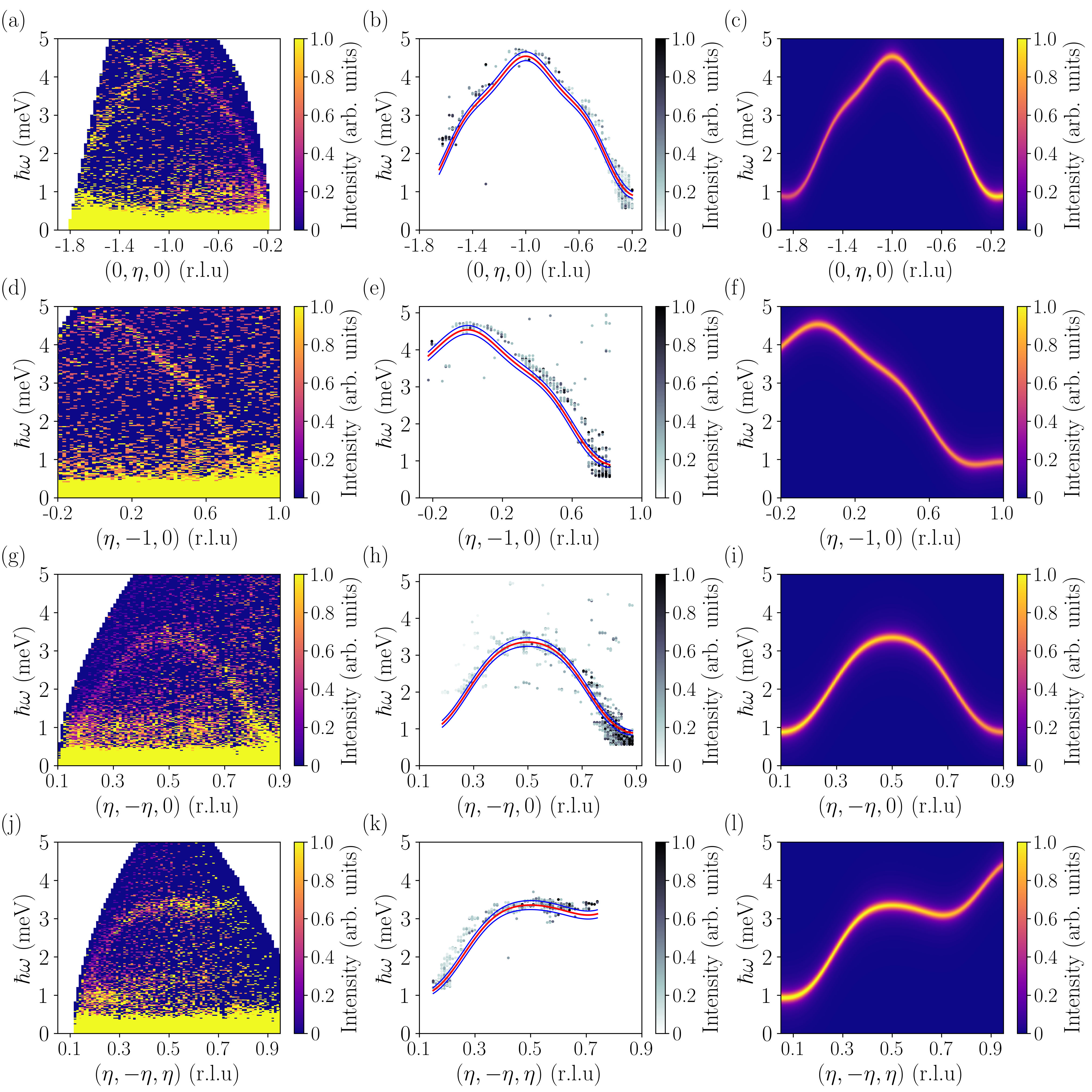} 
    \caption{Reciprocal space cuts used to fit the magnon model. The left-hand column shows the binned data. The central column shows the signal extracted data on which the fitting program operates. Overlaid on these is a red line showing the calculated dispersion relation and blue lines representing the upper and lower bound confidence intervals of the dispersion model, calculated using the parameters from Table 1 in the main manuscript. The right-hand column shows the $S(\mathbf{Q},\omega)$ calculation of the cut using the \textit{Su(n)ny} code\,\cite{zhang2021classical,dahlbom2022geometric}.}
    \label{fig: MultiBaltis1}
\end{figure}

\begin{figure}
    \centering
    \includegraphics[width=1\textwidth]{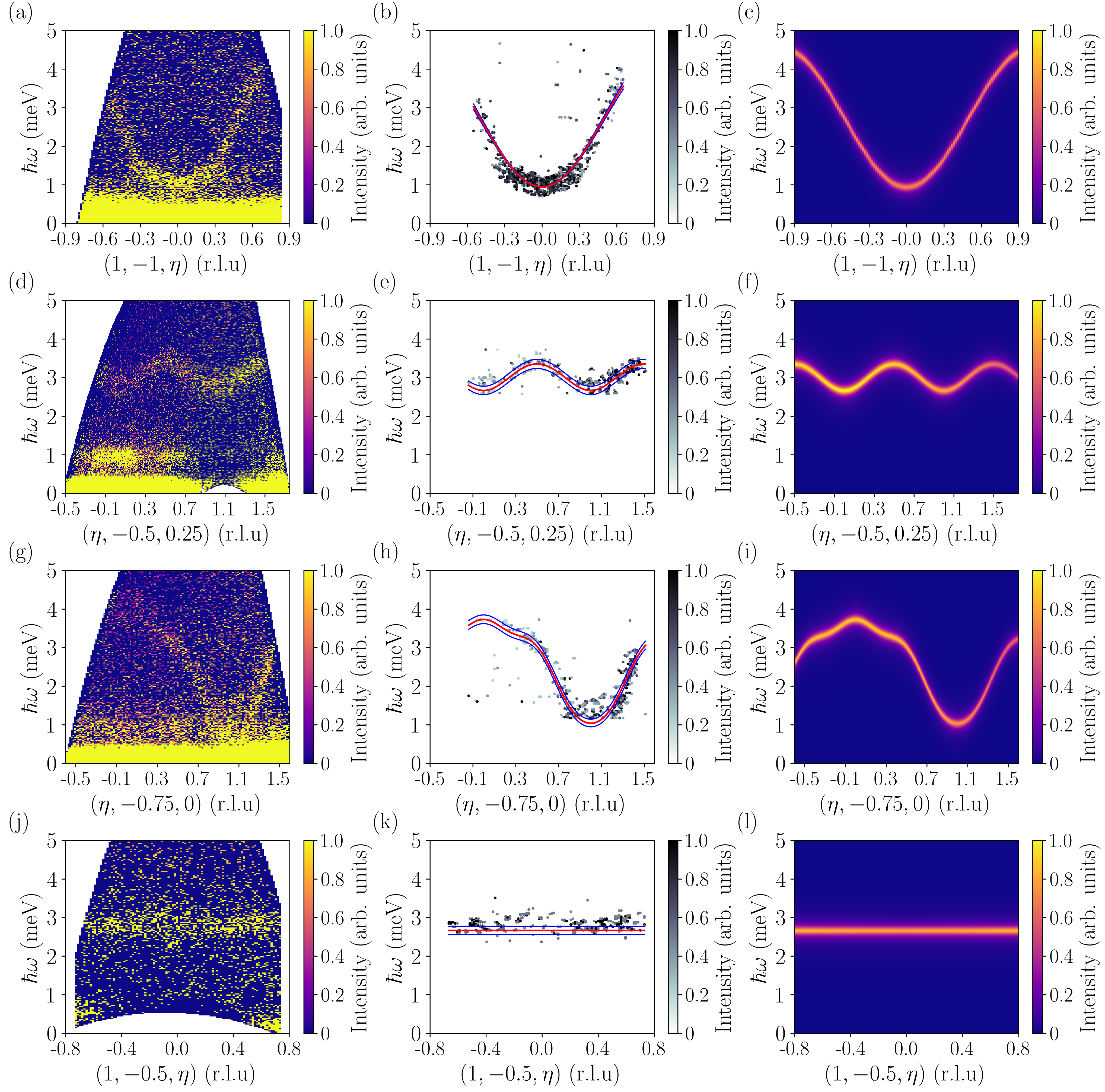} 
    \caption{Reciprocal space cuts used to fit the magnon model. The left-hand column shows the binned data. The central column shows the signal extracted data on which the fitting program operates. Overlaid on these is a red line showing the calculated dispersion relation and blue lines representing the upper and lower bound confidence intervals of the dispersion model, calculated using the parameters from Table 1 in the main manuscript. The right-hand column shows the $S(\mathbf{Q},\omega)$ calculation of the cut using the \textit{Su(n)ny} code\,\cite{zhang2021classical,dahlbom2022geometric}.}
    \label{fig: MultiBaltis2}
\end{figure}

\begin{figure}
    \centering
    \includegraphics[width=1\textwidth]{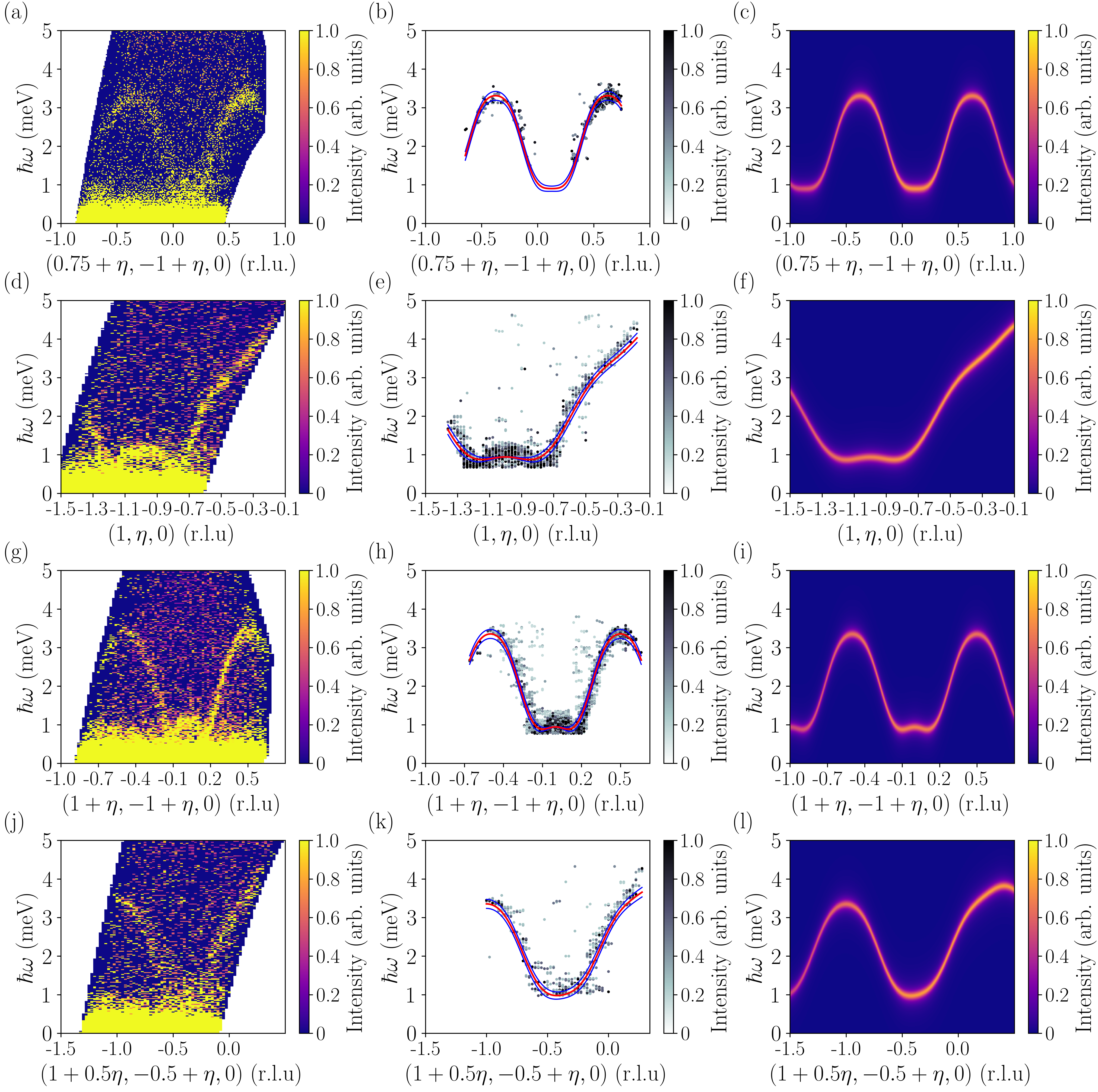} 
    \caption{Reciprocal space cuts used to fit the magnon model. The left-hand column shows the binned data. The central column shows the signal extracted data on which the fitting program operates. Overlaid on these is a red line showing the calculated dispersion relation and the blue lines representing the upper and lower bound confidence intervals of the dispersion model, calculated using the parameters from Table 1 in the main manuscript. The right-hand column shows the $S(\mathbf{Q},\omega)$ calculation of the cut using the \textit{Su(n)ny} code\,\cite{zhang2021classical,dahlbom2022geometric}.}
    \label{fig: MultiBaltis3}
\end{figure}

Figures \ref{fig: MultiBaltis1}, \ref{fig: MultiBaltis2} and \ref{fig: MultiBaltis3} show the complete set of reciprocal space cuts used in the fitting along with calculations of the model dispersion. The left-hand columns of these figures show the binned data along each reciprocal space path. The central columns show the corresponding data after the signal extraction has been applied. Overlaid on these is a red line representing the dispersion of the fitted $J_{1\rightarrow8}$ model and blue lines representing the upper and lower bounds of the calculated dispersion relation. These upper and lower bounds of the dispersion model are calculated through Monte-Carlo sampling of the error bounds shown in Table~1 in the main manuscript. The right-hand columns of Figs.~\ref{fig: MultiBaltis1}, \ref{fig: MultiBaltis2} and \ref{fig: MultiBaltis3} show the corresponding $S(\mathbf{Q},\omega)$ calculation of the $J_{1\rightarrow 8}$ model using the \textit{Su(n)ny} code\,\cite{zhang2021classical,dahlbom2022geometric}. It is worth noting that in the measurements, the intensity of the the magnon band appears stronger for high-Q scattering and weaker for low-Q scattering. This is the converse of what is conventionally expected for the intensity of a magnon band as the magnetic form factor decays with increasing $|\mathbf{Q}|$ which is consistent the the $S(\mathbf{Q},\omega)$ calculations\,\cite{boothroyd2020principles}. The reason the intensity of the measurement does not appear to follow the magnetic form factor is due to the absorption of the sample. In the low-Q regime, neutron scattering takes place in a forward scattering geometry, and therefore the neutron has to transmit through the sample, for which there is a high probability of the scattered neutron being absorbed. In the high-Q regime, the neutron scattering takes place in a backward scattering geometry and hence the neutron does not traverse as much sample, and hence the probability of the neutron being absorbed is much lower. 

The codes used to generate the cuts, perform the fittings and calculations are available upon request from the corresponding author.

\newpage
\section{The Method of Interaction Invariant Path Analysis}

In general, in the analysis of magnons from momentum-resolved spectra, various interactions will be highly correlated with each other. In fact, in many cases, along a given reciprocal-cut direction, different interactions may have $100\%$ correlations with one another. This occurs when the projection of the spin-wave theory for a given interaction along a chosen path is identical to that of another. As an example, consider $J_{1}$ (with $\boldsymbol{\delta}=[1,0,0]$) and $J_{9}$ (with $\boldsymbol{\delta}=[0,1,1]$) interactions~\footnote{Note that the inclusion of the symmetry equivalent $\boldsymbol{\delta}$'s are implicit here when referring to the $\boldsymbol{\delta}$ vector a given interaction.}. For reciprocal-space cuts within the $a^{*}$-$b^{*}$ plane it is clear from Eq.~3 in the main text that the dispersion curves for these interactions will project identically, and therefore it would only be possible to decorrelate these interactions by considering a cut that is active along $c^{*}$. Therefore, in deducing a Hamiltonian from momentum-resolved excitation spectra it is essential to analyze a judicious choice of reciprocal-space cuts so that the various interaction parameters may be refined as independently as possible. The \textit{Method of Interaction Invariant Path Analysis} provides a systematic way for choosing these reciprocal-space paths such that correlations between interaction parameters can be minimised. As shown directly in the main text (see Eq.~4), reciprocal space paths may be chosen such that a given interaction is non-dispersive by looking for solutions where the varying component of the dispersion relation is canceled out. This then allows other interactions which are dispersive along the cut to be refined.

\begin{figure}[htb]
    \centering
    \subfloat[$J_{2}$ invariant in the ($q_{x}, q_{y}$) plane.]{%
        \includegraphics[width=0.22\textwidth]{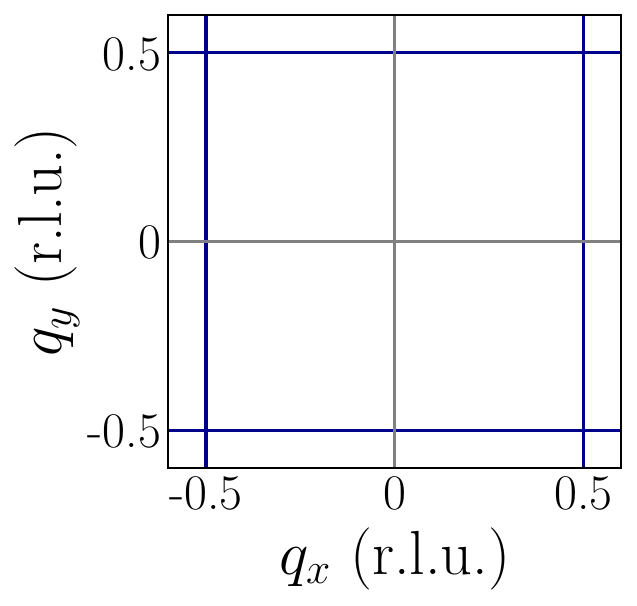}
        \label{fig:J2Inv_xy}
    }
    \hfill
    \subfloat[$J_{4}$ invariant in the ($q_{x}, q_{y}$) plane.]{%
        \includegraphics[width=0.22\textwidth]{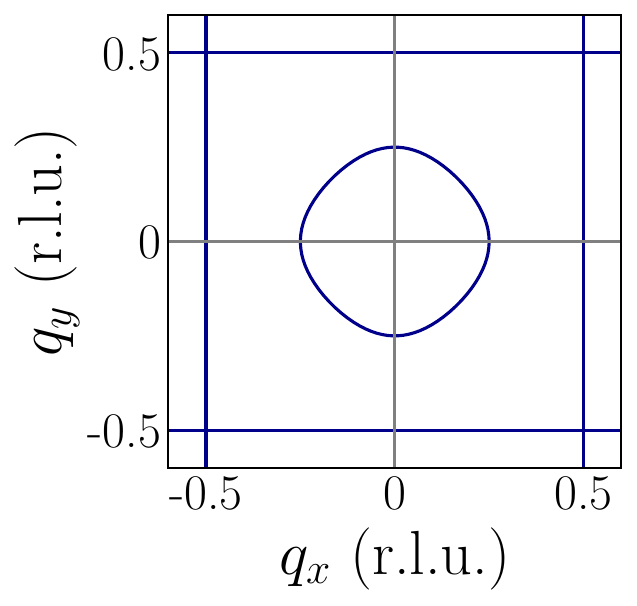}
        \label{fig:J4Inv_xy}
    }
    \hfill
    \subfloat[$J_{8}$ invariant in the ($q_{x}, q_{y}$) plane.]{%
        \includegraphics[width=0.22\textwidth]{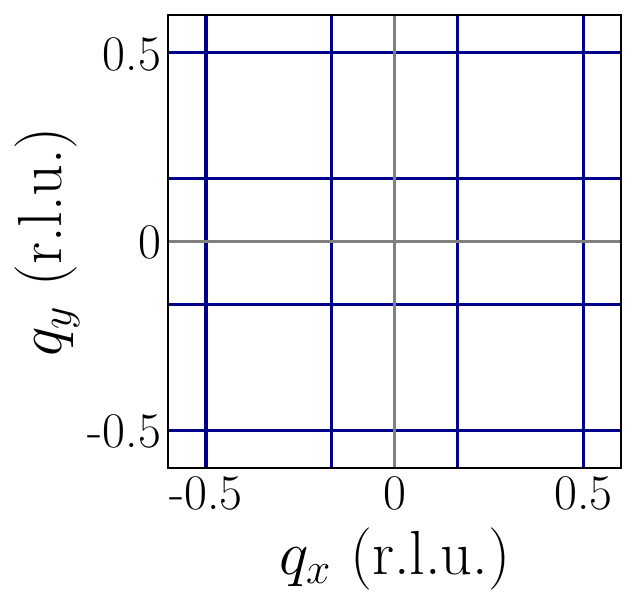}
        \label{fig:J8Inv_xy}
    }
    \hfill
    \subfloat[$J_{11}$ invariant in the ($q_{x}, q_{y}$) plane.]{%
        \includegraphics[width=0.22\textwidth]{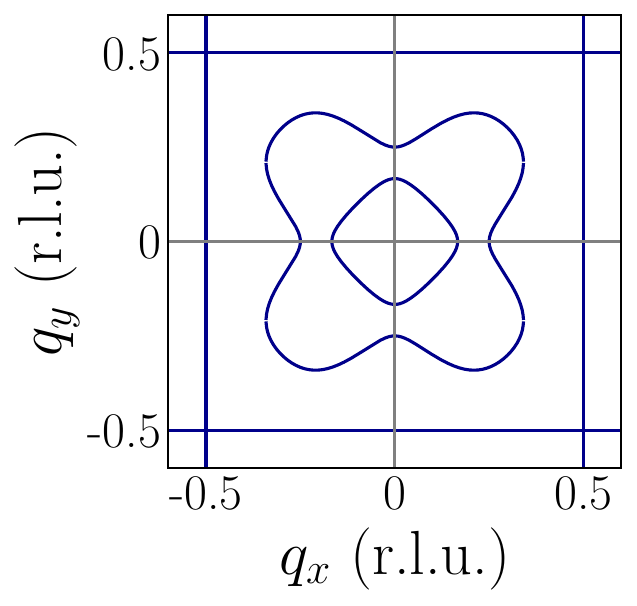}
        \label{fig:J11Inv_xy}
    }

    \vskip\baselineskip 

    \subfloat[$J_{2}$ invariant in the ($q_{x}, q_{z}$) plane.]{%
        \includegraphics[width=0.22\textwidth]{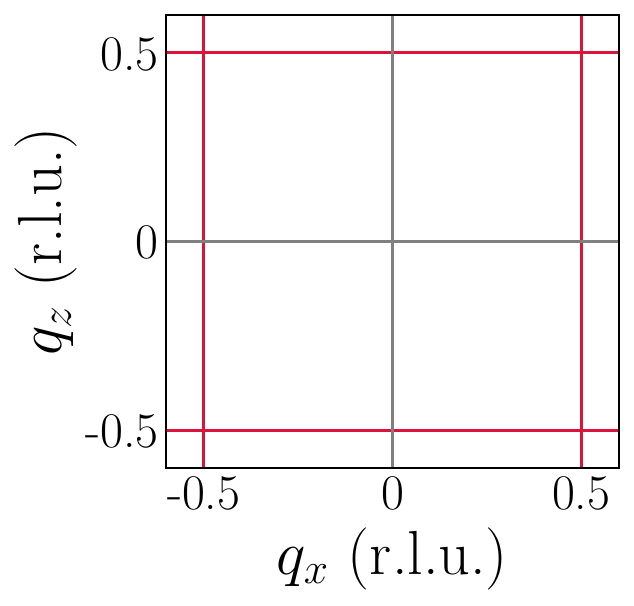}
        \label{fig:J2Inv_xzb}
    }
    \hfill
    \subfloat[$J_{4}$ invariant in the ($q_{x}, q_{z}$) plane.]{%
        \includegraphics[width=0.22\textwidth]{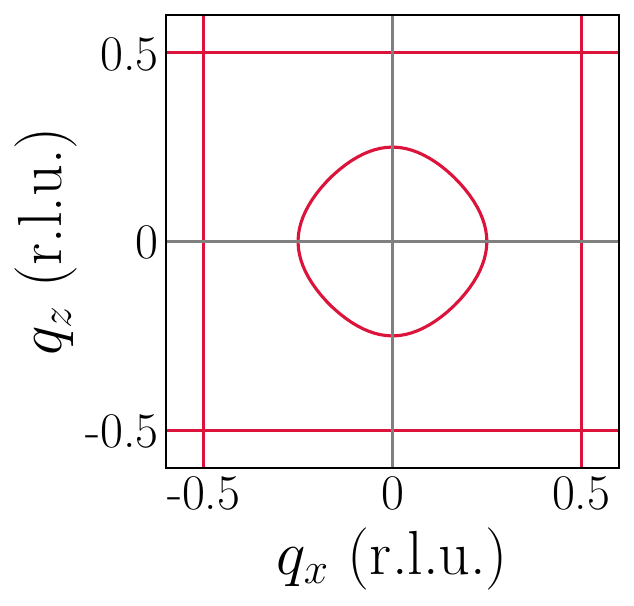}
        \label{fig:J4Inv_xzb}
    }
    \hfill
    \subfloat[$J_{8}$ invariant in the ($q_{x}, q_{z}$) plane.]{%
        \includegraphics[width=0.22\textwidth]{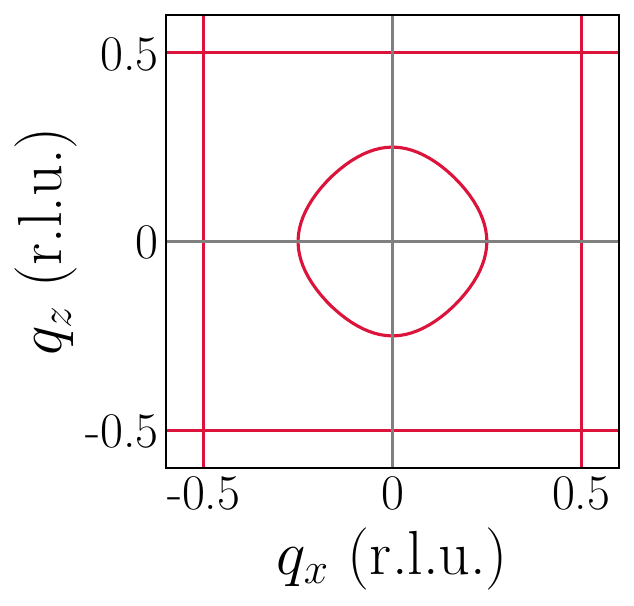}
        \label{fig:J8Inv_xzb}
    }
    \hfill
    \subfloat[$J_{11}$ invariant in the ($q_{x}, q_{z}$) plane.]{%
        \includegraphics[width=0.22\textwidth]{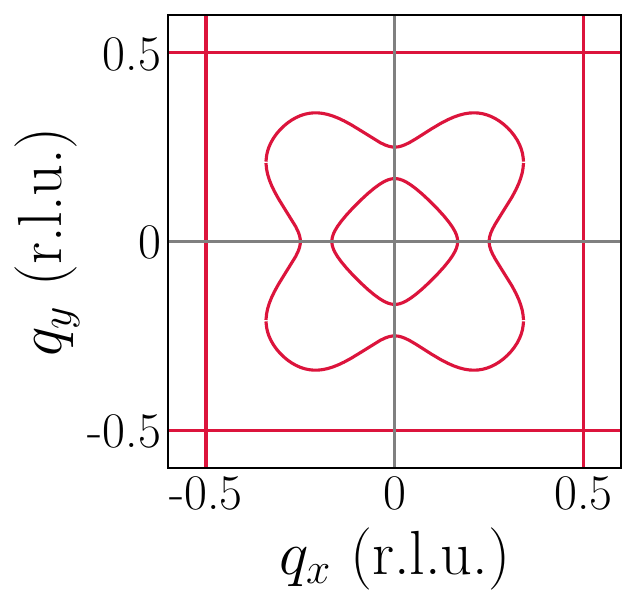}
        \label{fig:J11Inv_xzb}
    }
    \caption{Interaction invariants for the $J_{2}$, $J_{4}$, $J_{8}$, and $J_{11}$ interactions within the first Brillouin zone. The navy blue lines are for cuts in the $(q_{x},q_{y},q_{z}=0)$ plane. The red lines are for cuts within the $(q_{x},q_{y}=0,q_{z})$ plane.}
    \label{fig:Invariants}
\end{figure}

For the interactions examined, it has been found that each interaction has a unique set of invariant paths for distinguishibility. That is to say that quite frequently interactions may share certain invariants with one another; however, when considering two symmetry inequivalent reciprocal space planes, the set of invariants pertaining to a given interaction is unique. As an example, consider the $J_{2}$ ($\boldsymbol{\delta}=[1/2,1/2,1/2]$), $J_{4}$ ($\boldsymbol{\delta}=[3/2,1/2,1/2]$), $J_{8}$ $\boldsymbol({\delta}=[3/2,3/2,1/2]$) and $J_{11}$ ($\boldsymbol{\delta}=[5/2,1/2,1/2]$) interactions. The invariant paths for these interactions in the $q_{z}=0$ plane and the $q_{y}=0$ plane are shown in Fig.~\ref{fig:Invariants}. As can be seen, these interactions all share the linear $q_{x}$, $q_{y}$ or $q_{z}=1/2$ invariants. However, the $J_{4}$, $J_{8}$ and $J_{11}$ have additional invariant paths, some of which are non-linear loops, which make the set of invariants for a given interaction unique.

The form of $J_{2}$, $J_{4}$, $J_{8}$ and $J_{11}$ interactions are now provided. However, it should be noted that due to current software limitations, only linear invariant paths have been used within the analysis of this paper. The substitution $\iota=2\pi q_{x}$, $\kappa=2\pi q_{y}$ and $\varsigma = 2\pi q_{z} $ are made throughout and $n\in\mathbb{Z}$.

The interaction invariants for the $J_{2}$ interaction with $\boldsymbol{\delta}=[1/2,1/2,1/2]$ are given by,
\begin{equation}
(q_{x},q_{y},q_{z}=0)
\begin{cases}
\iota = 2\pi n + \pi, \\
\kappa = 2\pi n + \pi, \\
\end{cases}
\label{j2invariant1}
\end{equation}

\begin{equation}
(q_{x},q_{y}=0,q_{z})
\begin{cases}
\iota = 2\pi n + \pi, \\
\varsigma = 2\pi n + \pi. \\
\end{cases}
\label{j2invariant2}
\end{equation}

The interaction invariants for the $J_{4}$ interaction with $\boldsymbol{\delta}=[3/2,1/2,1/2]$ are given by,

\begin{equation}
(q_{x},q_{y},q_{z}=0)
\begin{cases}
\iota = 2\pi n + \pi, \\
\kappa = 2\pi n + \pi, \\
\kappa = 2\pi n \pm \arccos(1-\cos(\iota)),\\
\end{cases}
\label{j4invariant1}
\end{equation}

\begin{equation}
(q_{x},q_{y}=0,q_{z})
\begin{cases}
\iota = 2\pi n + \pi, \\
\varsigma = 2\pi n + \pi, \\
\varsigma = 2\pi n \pm \arccos(1-\cos(\iota)).\\
\end{cases}
\label{j4invariant2}
\end{equation}

The interaction invariants for the $J_{8}$ interaction with $\boldsymbol{\delta}=[3/2,3/2,1/2]$ are given by,

\begin{equation}
(q_{x},q_{y},q_{z}=0)
\begin{cases}
\iota = \frac{2\pi n}{3} + \pi,\\
\kappa = \frac{2\pi n}{3} + \pi,\\
\end{cases}
\label{j8invariant1}
\end{equation}

\begin{equation}
(q_{x},q_{y}=0,q_{z})
\begin{cases}
\varsigma = 2\pi n + \pi, \\
\iota = 2\pi n + \pi, \\
\varsigma = 2\pi n \pm \arccos(1-\cos(\iota)).\\
\end{cases}
\label{j8invariant2}
\end{equation}

The interaction invariants for the $J_{11}$ interaction with $\boldsymbol{\delta}=[5/2,1/2,1/2]$ are given by,

\begin{equation}
(q_{x},q_{y},q_{z}=0)
\begin{cases}
\iota = \pi,  \\
\kappa = \pi, \\
\kappa=\pm\arccos\left( \frac{\sqrt{5+\sqrt{8\cos(\iota)-8\cos(2\iota)+1}}}{2\sqrt{2}} \right),\\
\kappa=\pm\arccos\left( \frac{\sqrt{5-\sqrt{8\cos(\iota)-8\cos(2\iota)+1}}}{2\sqrt{2}} \right),\\
\end{cases}
\label{j11invariant1}
\end{equation}

\begin{equation}
(q_{x},q_{y}=0,q_{z})
\begin{cases}
\iota = \pi,  \\
\varsigma = \pi, \\
\varsigma=\pm\arccos\left( \frac{\sqrt{5+\sqrt{8\cos(\iota)-8\cos(2\iota)+1}}}{2\sqrt{2}} \right),\\
\varsigma=\pm\arccos\left( \frac{\sqrt{5-\sqrt{8\cos(\iota)-8\cos(2\iota)+1}}}{2\sqrt{2}} \right).\\
\end{cases}
\label{j11invariant2}
\end{equation}
\\

In the analysis of the magnon band in this study, linear $J_{2}$, $J_{4}$, $J_{8}$ invariants are used in Figs.~\ref{fig: MultiBaltis2}(d) and \ref{fig: MultiBaltis2}(j) and then $J_{5}$ and $J_{7}$ invariants in Fig.~\ref{fig: MultiBaltis3}(a) and finally Fig.~\ref{fig: MultiBaltis2}(g) is invariant against the $J_{3}$ and $J_{7}$ interactions~\footnote{The $J_{7}$ interaction is trivially invariant against any reciprocal path which is not active along $c^{*}$.}. Due to the limitations of the reciprocal-space coverage of the $90^{\circ}$ \textit{Horace} scan and the highly anisotropic absorption of the sample, linear invariants have not been used for the $J_{1}$ and $J_{6}$ interactions in this analysis. Irrespective of this, we have found that including the aforementioned invariants significantly informs the fitted model. 

We expect that this new methodology will find frequent application in the future analyzes of momentum-resolved excitation spectra, especially so once non-linear loop paths have been implemented in the cutting codes or indeed in determining which reciprocal paths to measure in a triple-axis spectrometer experiment.

\newpage
\section{Dipolar Interactions}

\begin{figure*}[h]
    \centering
    \includegraphics[width=0.5\textwidth]{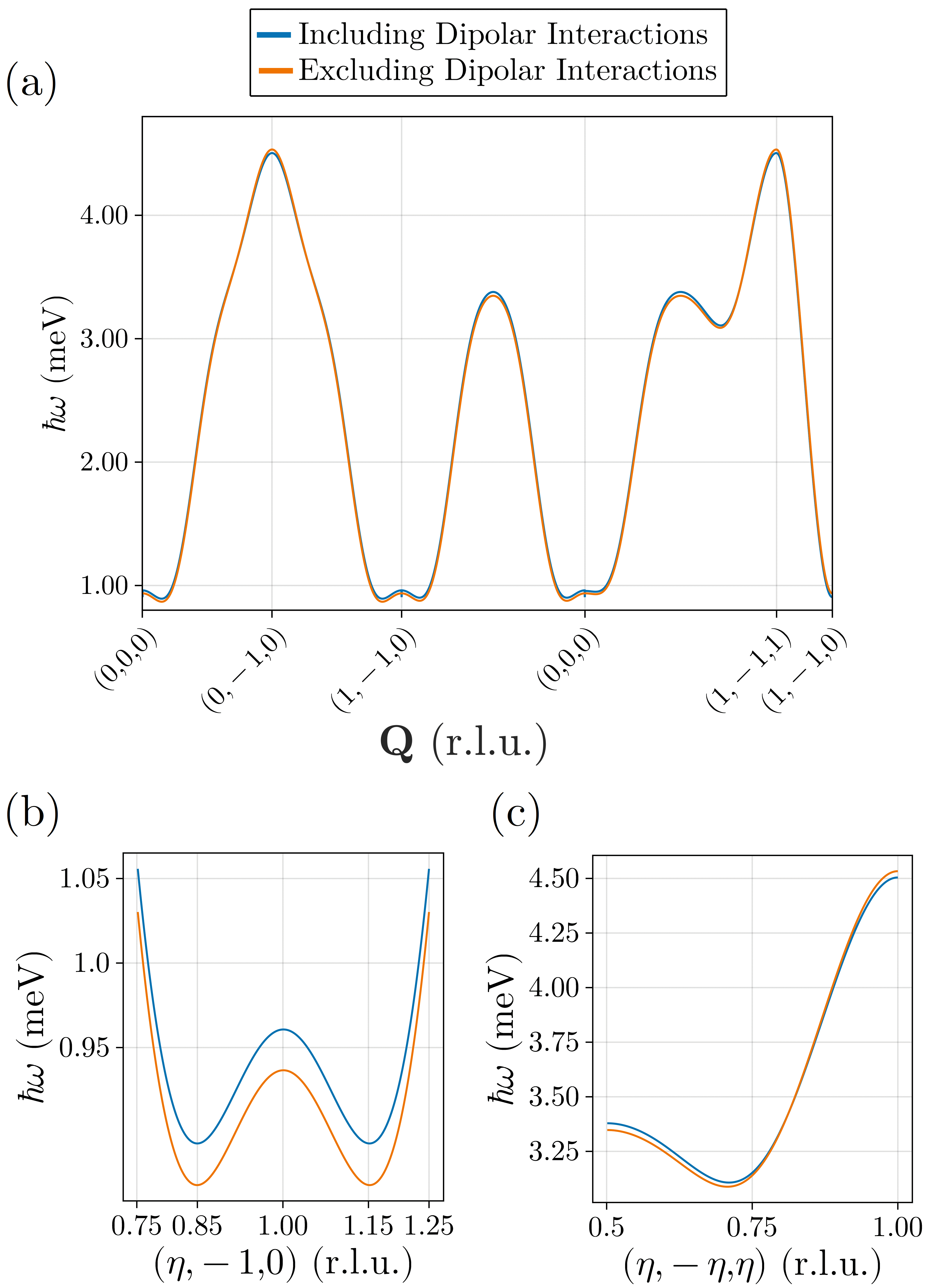}
    \caption{Model dispersion curves for \ce{GdRu2Si2}. The blue dispersion relation is the model including the dipolar interaction while the orange dispersion relation represents the original model presented in the paper \textit{i.e.} excluding the dipolar interactions. (a) An overview of the dispersion model across several different reciprocal paths. (b) and (c) are a zoomed in views of the dispersion relation along the $(\eta,-1,0)$ and $(\eta,-\eta,\eta)$ paths, respectively. As can be seen, the inclusion of the dipolar interaction has a small effect on the original dispersion curve (much finer than the resolution of the data) raising or lowering the dispersion relation by approximately $20~\si{\micro\eV}$. The dipolar interaction spin wave calculations were implemented in the \textit{Su(n)ny} code with Ewald summation~\cite{zhang2021classical,dahlbom2022geometric}.}
    \label{fig: triple}
\end{figure*}

\newpage
\section{Magnetometry with Curie-Weiss Fit}

\begin{figure*}[h]
    \centering
    \includegraphics[width=0.6\textwidth]{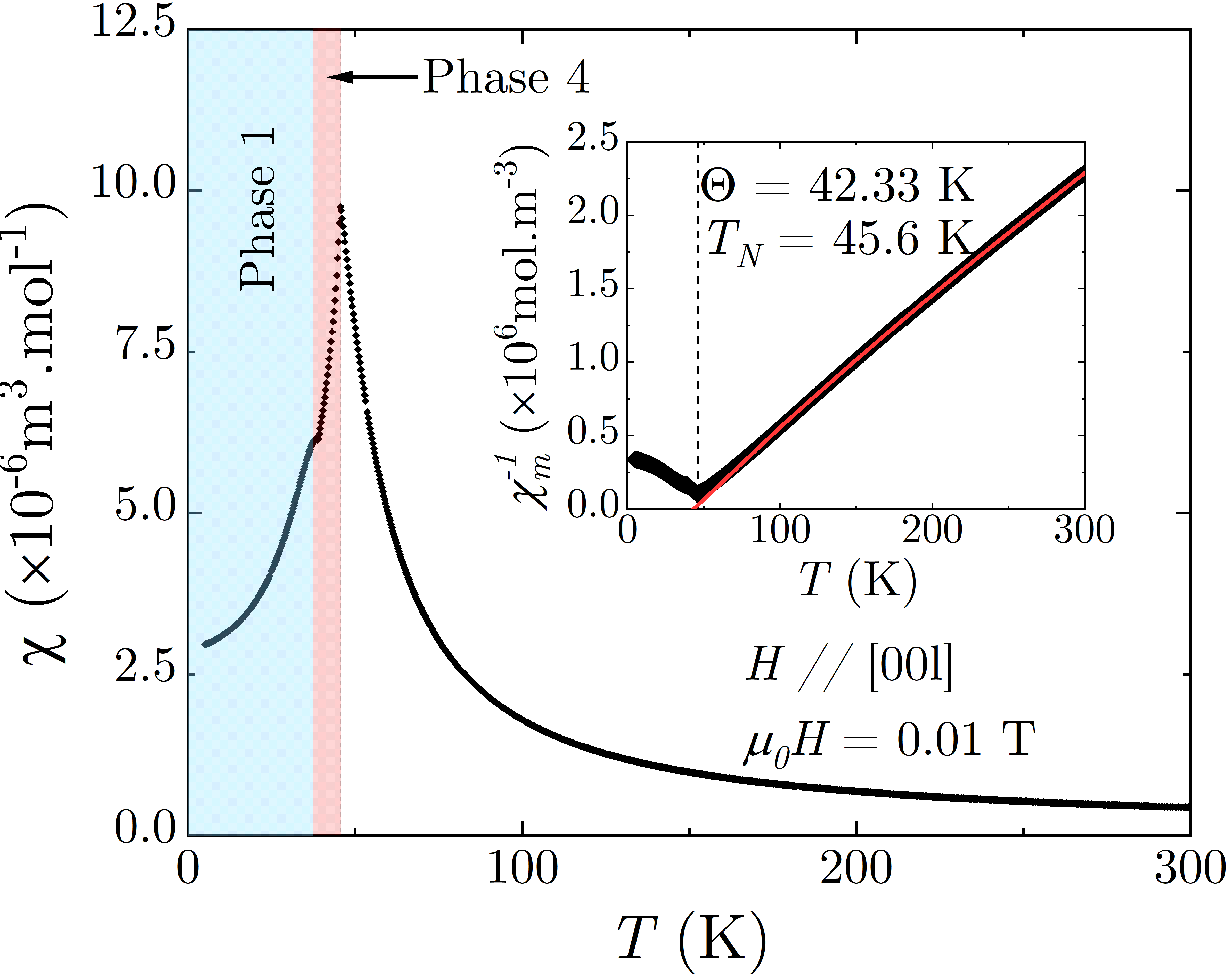}
    \caption{Magnetic susceptibility $\chi$ versus temperature $T$ for single crystal \ce{GdRu2Si2}. The inset shows $\chi^{-1}$ versus $T$ and a fit to a Curie-Weiss law (represented by the red line), with $\Theta=+42.33(4)\,\si{K}$. Measurements were made on a Quantum Design MPMS squid magnetometer with a magnetic field $\mu_{0}H=0.01\,\si{T}$ applied along the $c$~axis.}
    \label{fig: CW}
\end{figure*}

\newpage

\bibliography{Main.bib}